\documentclass[12pt,preprint]{aastex}

\usepackage{graphicx}
\usepackage{epstopdf}
\usepackage{amssymb,amsmath}

\usepackage{natbib}
\newcommand{\HII}{\hbox{H\,{\sc ii}}}
\newcommand{\HI}{\hbox{H\,{\sc i}}}

\newcommand{\msolar}{\mbox{\,M$_\odot$}\ }        % solar mass
\shorttitle{Cold Molecular Gas in NGC\,4151}
\shortauthors{Dumas et al.}

\begin{document}


\title{Cold Molecular Gas in the Inner Two Kiloparsec of NGC\,4151\footnote{Based on observations carried out with the IRAM Plateau de Bure Interferometer. IRAM is supported by INSU/CNRS (France), MPG (Germany) and IGN (Spain).}}


\author{G. Dumas}
\affil{Max-Planck-Institut f\"ur Astronomie, K\"onigstuhl 17, D-69117 Heidelberg, Germany}

\author{E. Schinnerer}
\affil{Max-Planck-Institut f\"ur Astronomie, K\"onigstuhl 17, D-69117 Heidelberg, Germany}

\author{C.G. Mundell}
\affil{Astrophysics Research Institute, Liverpool John Moores University, Twelve Quays House, Egerton Wharf, Birkenhead, CH41 1LD, UK}


\begin{abstract}

  We present the first spatially resolved spectroscopic imaging observations of the $^{12}$CO\,(1-0) line emission in the central 2.5\,kpc of the Seyfert 1 galaxy NGC\,4151, obtained with the IRAM Plateau de Bure Interferometer (PdBI). Most of the cold molecular gas is distributed along two curved gas lanes about 1\,kpc north and south of the active nucleus, coincident with the circumnuclear dust ring noted by previous authors. These CO arcs lie within the Inner Lindblad Resonance of the large scale oval bar and have kinematics consistent with those derived from neutral hydrogen observations of the disk and bar. Two additional gas clumps are detected that show non-circular motions -  one associated with the southern gas lane and one lying $\sim$600\,pc north of the nucleus. Closer to the nucleus, no cold molecular gas is detected in the central 300\,pc where abundant near-IR H$_2$ line emission arises.  This suggests that the H$_2$ line emission is not a good indicator for a cold gas reservoir in NGC\,4151 and that the H$_2$ is likely photo-excited by the AGN. The upper limit of the CO mass in the central 300\,pc is sufficient to support the AGN activity at its current level for $10^7$yrs.  
The total cold molecular mass detected by PdBI is $\rm 4.3\times10^7\,M_{\odot}$. Finally, 3\,mm continuum emission arising from the location of the AGN is detected  with a flux of $S_{3mm}~\sim~14$~mJy and appears to be unresolved at an angular resolution of 2\farcs8 ($\sim$180 pc).
\end{abstract}


%% subject header.  Objects should be in the appropriate "individual"
%% headers (e.g. quasars: individual, stars: individual, etc.) with the
%% additional provision that the total number of headers, including each
%% individual object, not exceed six.  The \objectname{} macro, and its
%% alias \object{}, is used to mark each object.  The macro takes the object
%% name as its primary argument.  This name will appear in the paper
%% and serve as the link's anchor in the electronic edition if the name
%% is recognized by the data centers.  The macro also takes an optional
%% argument in parentheses in cases where the data center identification
%% differs from what is to be printed in the paper.


\section{Introduction}

Nuclear activity in galaxies is thought to be driven by the release of 
gravitational potential energy from material accreted by a central 
supermassive black hole. In this scenario, fuel transportation mechanisms 
in the nuclear regions must be efficient on time scales comparable to 
that of the nuclear activity of the order of  $10^7$-$10^8$\,yrs \citep[e.g.][]{martini_04, marconi_04}.
Stellar bars are well known to drive gas from the outer large-scale
disk to the central kiloparsec \citep{sakamoto_99, kartik_05} and the
bar-within-bar scenario \citep{shlosman_89} was proposed as a mechanism 
to deliver gas closer to the central black hole. Recently, other gravitational
perturbations, such as m=1 modes or gas density waves, have been
suggested to play a major role in the transport of gas to the most central
regions \citep[e.g.][]{englmaier_00,emsellem_01,NUGA_1st_03}.

Molecular gas dominates the interstellar medium (ISM) in the
centers of nearby spiral galaxies, therefore CO lines are the best
tracers for the nuclear gas distribution and dynamics. High resolution
CO observations of the central kpc regions of nearby active galaxies
provide insights on the ISM properties in these regions and the
processes that could move gas inwards. The recently completed NUGA
survey with the IRAM mm-interferometer PdBI (Plateau de Bure
interferometer) \citep{NUGA_1st_03} observed 12 Seyfert and LINER
galaxies with different nuclear and host properties and showed that
molecular gas is present in the centers, however with diverse
morphology and kinematics. This already suggests that more than one
single mechanism acting in the central kpc might be responsible for
fueling the black hole \citep{NUGA_1st_03, garcia-burillo_05,
  combes_04,krips_05,boone_07,hunt_08,lindt_08,casasola_08,haan_09,combes_09}.
Thus, the mechanism for transporting material from the central few hundred pc to the 
SMBH in order to induce and sustain the nuclear activity remains debated.
Observations of molecular gas in nearby AGN can begin to bridge this gap, 
connecting large and small spatial scales and examining the next link in the fueling chain. 
In this context, the nearby spiral galaxy NGC\,4151 hosting a Seyfert 1 nucleus is an ideal target.

NGC\,4151 is one of the best studied Active Galactic Nuclei (AGN)
\citep[see review by][]{ulrich_00} and situated at a distance of
13.3Mpc \citep[1\arcsec\ = 65pc,][]{mundell_03}. It is a barred spiral galaxy, 
classified as (R')SAB(rs)ab
\citep{RC3}, and seen nearly face-on (i=21\degr). In addition to a
remarkably large stellar bar (3\arcmin$\times$2\arcmin\ $\rm \approx
11.7\,kpc\times7.8\,kpc$), two faint spiral arms extend out to
a radius of $\sim$6\arcmin\ (23.4\,kpc). It has been extensively observed at all
wavelengths and on various spatial scales. In particular neutral
hydrogen (\HI) traces the large-scale stellar bar and the outer spiral arms. 
On kpc scales, \cite{vila-vilaro_95} have detected two red arc-like features,
possibly produced by dust extinction of the background stellar
continuum, which delineate a circumnuclear elliptical ring (hereafter:
central dust ellipse) with a semi-major axis of $\sim$18''
($\sim$1.2\,kpc). In addition a large Extended Narrow Line Region (ENLR) extends up to
20\arcsec\ (1.3kpc) and is consistent with ambient galactic gas being
photo-ionized by a cone of nuclear UV radiation \citep{penston_90}. Evidence 
for radial outflow have been observed and modelled in the NLR of NGC4151 \citep{Heckman_83, schulz_90, Hutchings_98, Kaiser_00, crenshaw_00,das_05,storchi-bergmann_10}.
The bright Seyfert nucleus of type
1.5 displays rapid optical variability, and a small $\sim$ 600 pc
long radio jet at a position angle (PA) of $\sim$ 80\degr
\citep{pedlar_93}. On scales of about 1\arcsec\ to 2\arcsec,
optical CCD imaging revealed evidence for obscuration in the form of a
reddened band of enhanced extinction (or low ionization) crossing the
central region, and aligned approximately perpendicular to the ENLR
and radio jet \citep{perez_89, terlevich_91}. In addition, there is also evidence 
of neutral gas and dust within $\sim$10\,pc of the
nucleus \citep{mundell_95,mundell_99_2}.  A summary of the properties of NGC4151 is provided
in Tab.~\ref{properties}.

Here we present the first mm-interferometric $^{12}$CO(1-0)
observations of the central 2\,kpc of NGC4151, encompassing the dust
ellipse, obtained with PdBI. These data provide a link between
the large-scale bar and the (putative) torus seen in \HI\ in absorption
\citep{mundell_03}. We describe the observations and data
reduction in Sect.~\ref{data_reduc}. The resulting CO maps and
molecular gas properties are presented in Sect.~\ref{results}. The
relation between the molecular gas and the large-scale properties of
NGC4151 is analyzed in Sect.~\ref{large_scale}. In Sect.~\ref{COvsH2},
cold  and warm molecular gas properties in the innermost region are compared and we discuss
our results in the context of AGN fueling in Sect.~\ref{fueling}. 
Finally we conclude in Sect.\ref{conclusion}.

\section{Observations and Data Reduction}
\label{data_reduc}

The $^{12}$CO(1-0) emission line at 3mm was observed with IRAM/PdBI
using 6 antennas in C and D configuration on December
26 and 27, 2008 and April 9 and 19, 2009, respectively, providing baselines from 
 24\,m to 176\,m.  The calibration and mapping
were done with the standard IRAM GILDAS software packages CLIC and
MAPPING \citep{MAPPING}. The phase center of the observation was at 
RA$=12^h10^m32.^s579$ and DEC$=+39\degr24\arcmin20\farcs63$ in the
J2000.0 coordinate system, all velocities are observed relative to
$v_{sys}=995$~km\,s$^{-1}$. The quasars 3C273, 3C84, 3C345, 1144+402,
1308+326, 0923+392, 0528+134 and 1156+295 were observed as flux
calibrators. During all observations, 1144+402 was used as phase
calibrator and observed every 24min.

We separately calibrated each day of observations using the
corresponding phase and flux calibrators. All observations were then
merged into a single dataset. The correlator was set to a spectral
resolution of 1.25MHz per channel, corresponding to a velocity
resolution of 3.25 km s$^{-1}$. We binned by two channels for a final
spectral resolution of 6.5 km s$^{-1}$. In addition, a pure continuum
map was created by averaging all channels with no line emission. In
order to obtain a data cube containing only emission line information,
the continuum was subtracted from the line data in the uv-plane using
the task UV\_SUBTRACT in GILDAS. The continuum and emission line only
data were cleaned separately, using uniform and natural weighting (see
Tab. \ref{beams}). All cleaned data have a pixel size of 0.5\arcsec,
with an image size of 256$\times$256 pixels .

\begin{figure}
\begin{center}
\includegraphics[width=10cm, angle=-90]{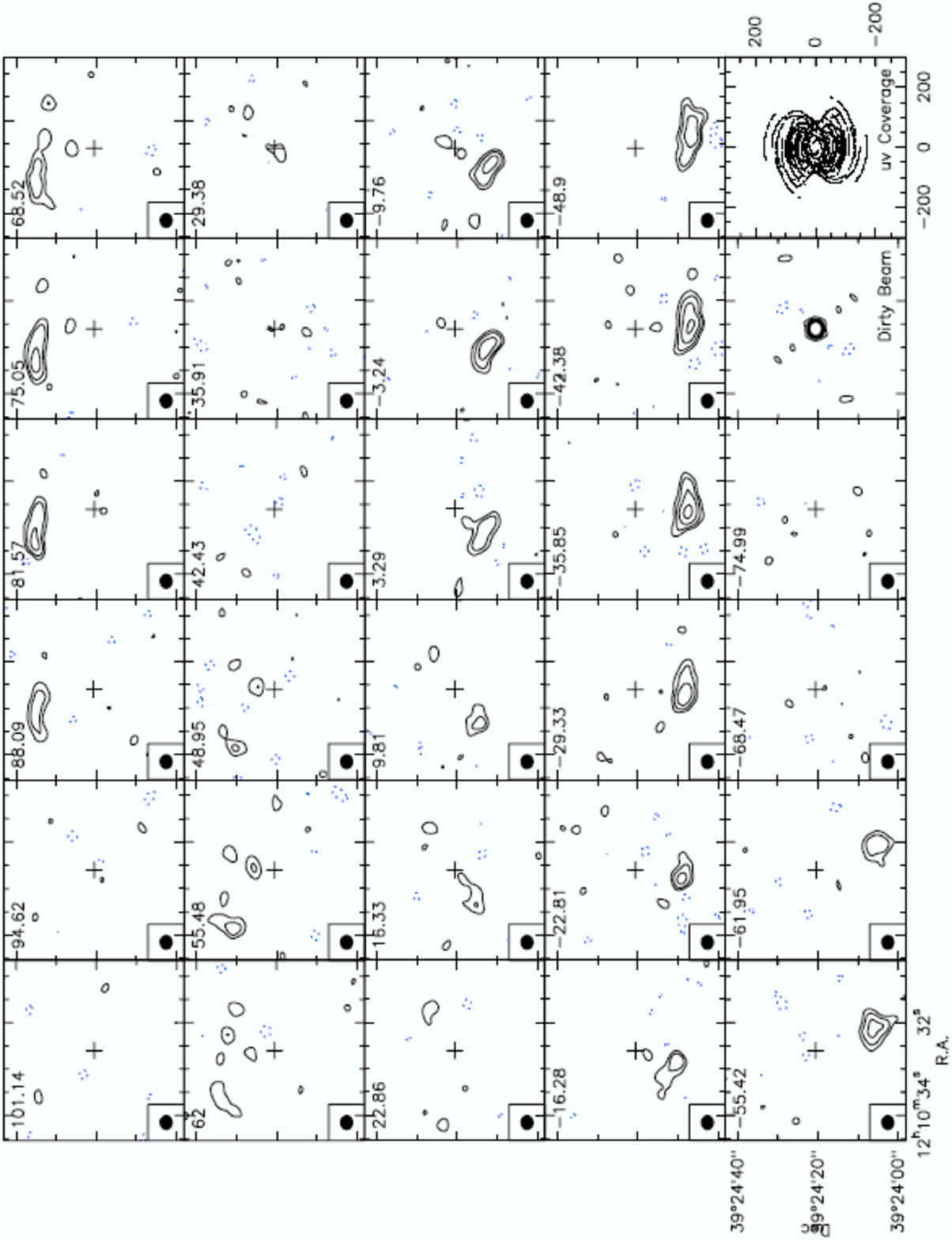} 
\end{center}
\caption{Channel maps of the naturally weighted CO(1-0) line emission
  in NGC4151. The channels are 6.5 km s$^{-1}$ wide and contours are
  plotted at -6, -3, 3, 6, 12 and 24$\sigma$ with 1$\sigma$ = 2\,mJy\,beam$^{-1}$. The velocity marking in the top left corner of
  each panel is relative to the observed central velocity of
  $v_{sys}=995$ km\,s$^{-1}$. The CLEAN beam of
  3.44\arcsec$\times$2.97\arcsec\ is shown in the bottom left corner of
  each panel. The cross marks the phase center of the observation. 
The dirty beam and the $uv$ coverage are presented in
  the last two panels.}
\label{channel}
\end{figure}

The channel maps of the naturally weighted CO(1-0) line are presented
in Fig.~\ref{channel} along with the dirty beam and the $uv$ coverage. All
data are presented without primary-beam correction applied. At the
observed frequency, the typical FWHM of the PdBI primary beam is
44\arcsec. Hereafter, we present the continuum emission at the
uniform resolution (2.83\arcsec$\times$2.16\arcsec ) and the results
from the naturally weighted emission line cube at 6.5~km s$^{-1}$
velocity resolution, which has a lower spatial resolution (beam of
3.44\arcsec$\times$2.96\arcsec\ or 224pc$\times$192pc) but a better
sensitivity, with a noise level of 2~mJy\,beam$^{-1}$, than
the uniformly weighted cube (for which the rms is 2.8~mJy\,beam$^{-1}$).

\section{Continuum, Molecular Gas Distribution and Kinematics}
\label{results}

Here we present the properties of the mm continuum and
CO(1-0) line emission in the central two kpc of NGC4151, as 
observed with PdBI.

\subsection{Millimeter continuum emission}

\begin{figure}
\begin{center}
\includegraphics[width=8cm]{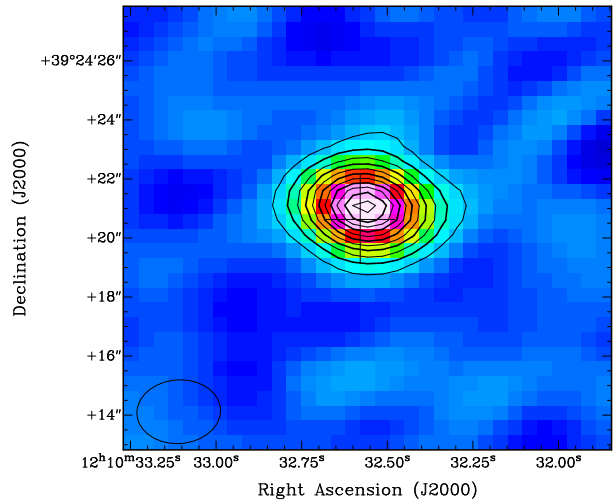} 
\end{center}\caption{3mm continuum map of NGC4151 using uniform
  weighting. The beam size of 2.8\arcsec$\times$2.2\arcsec\ is
  indicated in the bottom-left corner. The contour levels run from 1
  to 12~mJy beam$^{-1}$ with a step of 1~mJy beam$^{-1}$ (3.5$\sigma$
  to 43$\sigma$ in steps of 3.5$\sigma$).The cross marks the
    phase center of the observations, while the peak of the continuum
    coincides with the location of the AGN, within our astrometric uncertainty.}
\label{cont}
\end{figure}

The cleaned millimeter continuum map derived with uniform weighting is
shown in Fig.~\ref{cont}. The beam size is
2.8\arcsec$\times$2.2\arcsec\ and the rms noise is 0.3~mJy\,beam
$^{-1}$. The total integrated flux density is
14~mJy and the peak flux is 12~mJy\,beam$^{-1}$.
The millimeter continuum peaks at RA$=12^h10^m32.^s56$ and
DEC$=+39\degr24\arcmin21\farcs06$ (J2000.0) corresponding to the AGN
location \citep{mundell_03}, with an offset of 0.25\arcsec, consistent with our astrometric 
uncertainty of about 0.4\arcsec . At this resolution it is barely resolved:~a two-dimensional Gaussian fit
gives a convolved source size of 3.2\arcsec$\times$2.3\arcsec\ with a
PA of 93$\degr$. While the
elongation of the continuum is similar to the shape of the clean beam,
this suggests a small extent in east-west direction roughly consistent
with the orientation of the radio jet that is elongated along a PA of
77\degr\ \citep{mundell_95}. 

\subsection{CO line emission}

We used the task MOMENTS in GIPSY to construct the intensity, velocity
and velocity dispersion maps of the CO(1-0) line emission, using a
flux clipping level of 3$\sigma$=6~mJy\,beam$^{-1}$ per channel and 
considering emission only as real if it appears at
least in two consecutive channels. The width of 2 channels is 13\,km\,s$^{-1}$, 
smaller than the typical line width (20\,km\,s$^{-1}$) seen
throughout the data cube. The three moment maps of the naturally
weighted data cube are shown in Fig.~\ref{moments}.

\subsubsection{Molecular gas morphology and masses}

The integrated intensity map of the CO line emission (Fig.~\ref{moments}, 
top panel) shows a fairly regular geometry. About
15\arcsec\ north and south of the nucleus the molecular gas forms two
gas lanes that run roughly in east-west direction over a length of
16\arcsec\ (north) and 20\arcsec\ (south). The southern lane is about
twice as bright than the northern lane. In addition to these extended
structures, several fainter and more compact clumps are detected. All
major components are indicated in Fig.~\ref{components}. The most
prominent clump is closest to the nucleus (labeled as central clump in
Fig.~\ref{components}) at a distance of 9\arcsec\ ($\approx$600pc) to
the North. Four additional smaller clumps are present, three near
the northern lane and the last one at about 25\arcsec\ west (western
clump on Fig~\ref{components}) of the center of NGC4151.

 We measured the CO fluxes  within different components marked in Fig.~\ref{components} 
using the tasks BLOT (to define the regions) and FLUX
in GIPSY. All the fluxes presented here are corrected for primary beam attenuation. We then converted the CO(1-0) flux into molecular gas mass
for each component, using a
standard Galactic CO-to-H$_2$ conversion factor of
$X_{CO}=2\times10^{20}$cm$^{-2}$[K km s$^{-1}$]$^{-1}$ \citep{solomon_91}
 and applying a mass correction for helium of 36\% . The derived CO fluxes
and  molecular gas masses are listed in Tab.~\ref{mass}. In our entire
map, the total flux recovered is 14.7\,Jy\,km\,s$^{-1}$ and the corresponding cold molecular gas mass is about 4.3$\times$10$^7\,M_{\odot}$, with 60\%\
of this mass being contained in the southern lane.  Our PdBI data recover more flux than the JCMT single dish measurements \citep{rigopoulou_97}. They measure a total CO(2-1) flux of I$_{CO}=1.56$\,K\,km\,s$^{-1}$, corresponding to S$_{CO(1-0)}=11$\,Jy\,km\,s$^{-1}$ in integrated flux, assuming optically
 thick molecular gas, in the inner 23\arcsec\ ($\sim$1.5\,kpc). The total H$_2$ mass derived from the single dish data is then M(H$_2$)$=2.4\times10^7\,M_{\odot}$ which corresponds to a total molecular mass of M$_{mol}=3.2\times10^7\,M_{\odot}$, after applying the mass correction for helium. 
 Therefore, the JCMT data missed about 25\% of flux coming from the outer edge of the CO  gas lanes.

Interestingly, we detect no CO emission in the central 3\arcsec ,
where warm $H_2$ emission is observed via its NIR emission lines \citep{fernandez_99, storchi-bergmann_09_1}.
 We discuss the discrepancy between the CO and NIR H$_2$ observations in the central 
regions in Sect.~\ref{COvsH2}. 

\subsubsection{CO kinematics}
\label{COkin}
\begin{figure}
\begin{center}
\includegraphics[width=10cm]{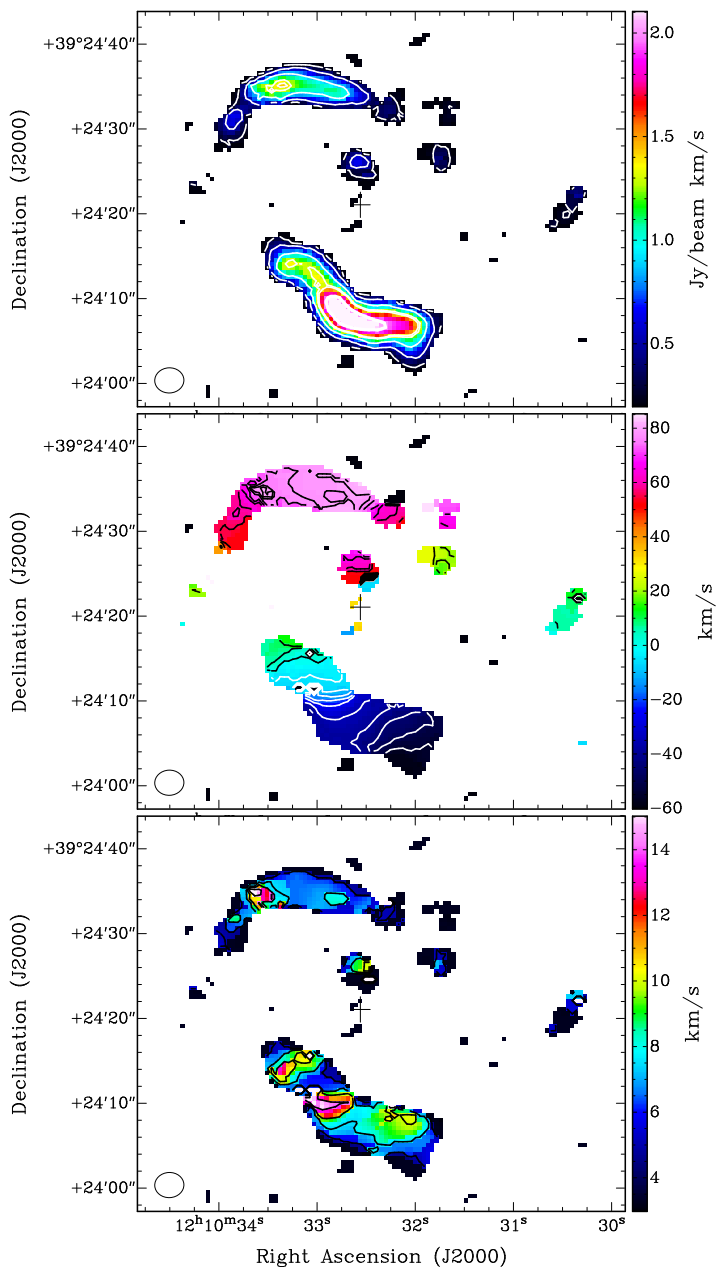} 
\end{center}
\caption{Moment maps of the CO(1-0) line emission in NGC\,4151:
  integrated intensity (top), velocity field (middle) and velocity
  dispersion (bottom). The contour levels are 0.25, 0.5, 0.9, 1.3, 1.5 and
2\, Jy\,beam$^{-1}$\,km\,s$^{-1}$  in the intensity map; from -50\,km\,s$^{-1}$ 
  to 80\,km\,s$^{-1}$ in steps of 5\,km\,s$^{-1}$ for the
  velocity field and 5, 7.5 10 and 14\,km\,s$^{-1}$ in the dispersion
  map. The beam size (3.4\arcsec$\times$3.0\arcsec) is shown in the
  bottom left corner of each panel and the cross marks the location of
  the AGN as given by its 21cm radio continuum \citep{mundell_03}.}
\label{moments} 
\end{figure}

\begin{figure}
\begin{center}
\includegraphics[width=8cm]{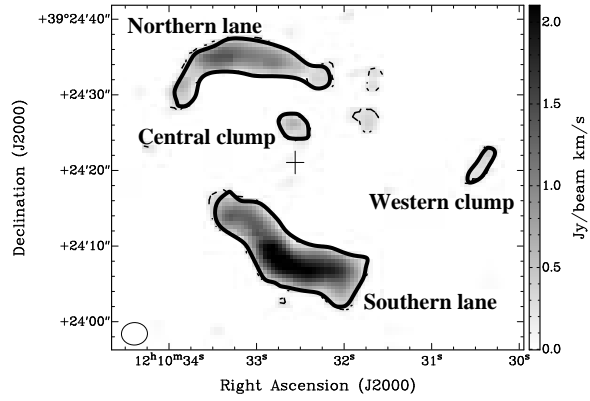} 
\end{center}\caption{The
  individual components of the CO(1-0) distribution discussed in the
  text are indicate in the intensity map. The contour marks the 0.21
  Jy\,beam$^{-1}$\,km\,s$^{-1}$ level and roughly outlines the
  areas used to measure the CO(1-0) line flux for each component (see
  Table~\ref{mass}).}
\label{components} 
\end{figure}

\begin{figure}
\begin{center}
\includegraphics[width=\textwidth]{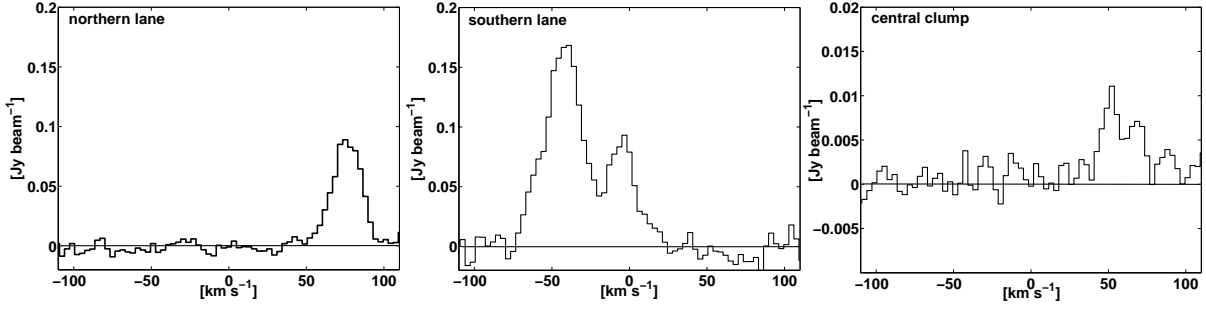} 
\end{center}\caption{Integrated CO spectra of the
  three major components of the CO distribution: northern gas lane ({\it
    left}), southern gas lane ({\it middle}), and  central clump ({\it
    right}). For all spectra, the x-axis is relative to the
  systemic velocity of v$_{sys}\,=\,995\,$km\,s$^{-1}$ (in km\,s$^{-1}$) and the y-axis is the integrated
  flux density in Jy\,beam$^{-1}$. Note the second component in
  the profile of the southern gas lane that is centered at 0\,km\,s$^{-1}$
  and mostly arising from the northeastern tip of this lane.}
\label{spectra_mean} 
\end{figure}

\begin{figure}
\begin{center}
\includegraphics[width=\textwidth]{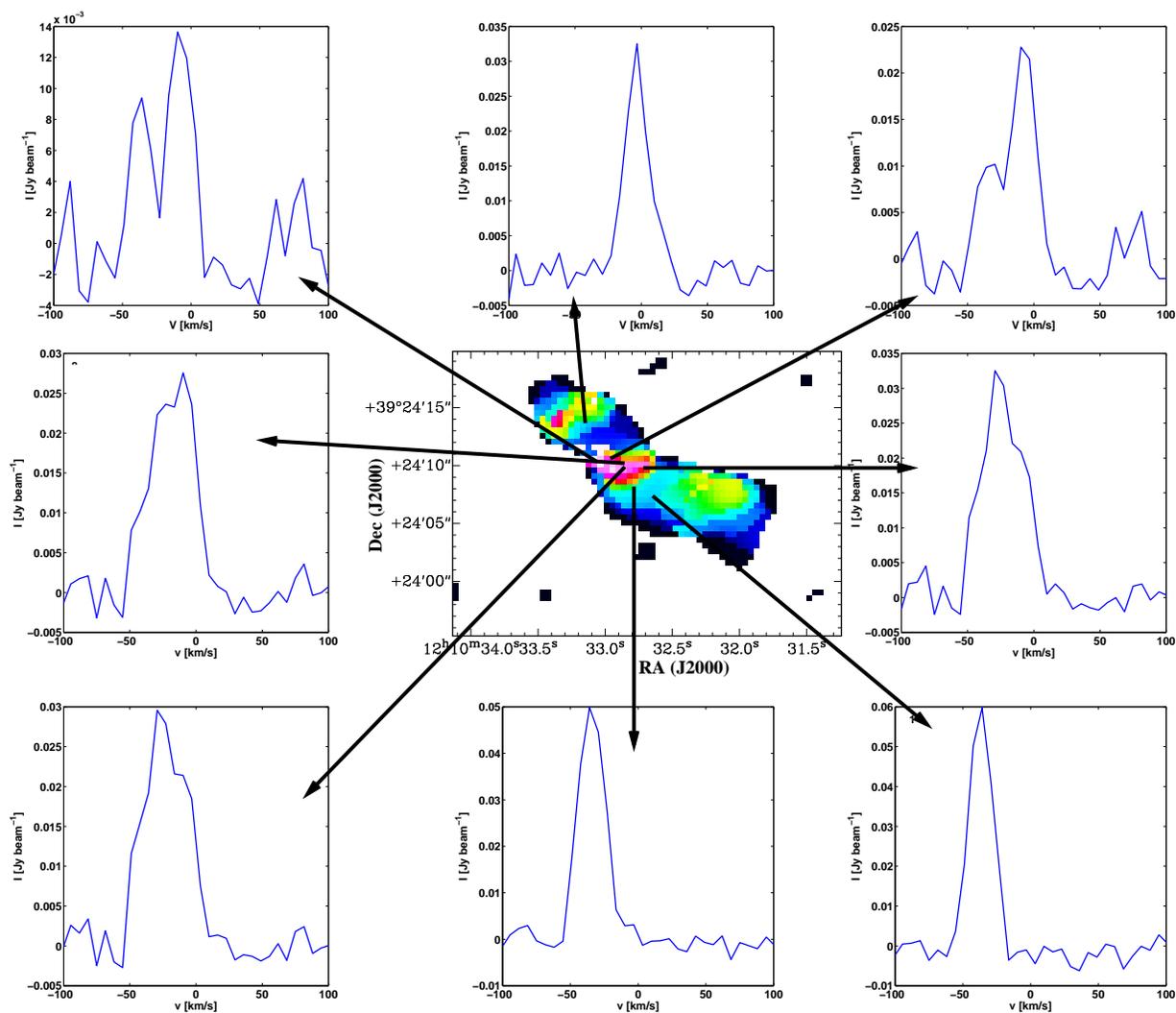} 
\end{center}\caption{CO emission line profiles from different regions of the southern lane. The central panel shows the velocity dispersion accross the southern lane, between 3 and 15\,km\,s$^{-1}$ from blue to white colors. The spectra have been extracted from single pixels of size 0.5\arcsec$\times$0.5\arcsec, the locations of these pixels are indicated by the arrows.}
\label{double_peaks} 
\end{figure}

The CO(1-0) velocity field and the velocity dispersion map are
presented in the middle and bottom panels of Fig.~\ref{moments}. The velocity field in the extended southern 
and northern lanes shows an ordered gradient along the lanes with values similar to those derived from HI observations \citep{asif_98,mundell_99_2} and
an  average velocity dispersion of approximately 7\,km\,s$^{-1}$.  Fig.~\ref{spectra_mean} shows the
integrated spectra of the three main components: the northern and
southern gas lanes as well as the central clump. The integrated spectrum of the northern lane is centered on 80\,km\,s$^{-1}$ and the central clump at 55\,km\,$s^{-1}$. The spectrum of the southern lane shows the presence of two kinematic components  originating from each half of the lane - one centered on 0\,km\,s$^{-1}$, corresponding to the eastern part of the lane where it curves to the North, and the other centered on $-40$\,km\,s$^{-1}$ arising from its most prominent western part - separated by a region in the centre of the lane that shows an increased velocity dispersion, close to the kink point.   Fig.~\ref{double_peaks} shows a selection of spectra taken close to this high velocity dispersion region. Asymmetry in the line profiles is clear, suggesting multiple kinematic components, but this region is spatially only marginally resolved while a second line component can be easily identified in several spectral. This boundary region between the two halves of the gas lane therefore may represent a region in which gas clouds with distinct velocities overlap along the line of sight or a genuinely mixed, disturbed zone.
 Larger velocity dispersion in the east part of the northern lane
may also indicate that the eastern part of this lane is a distinct cloud and although the line profiles in this region deviate from a simple Gaussian shape they show 
no clear evidence of two components in this region.

\begin{figure}
\begin{center}
\includegraphics[width=\textwidth]{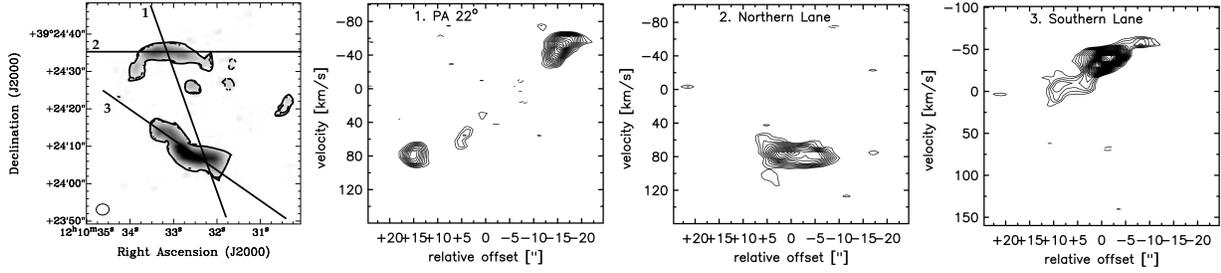} 
\end{center}\caption{CO $pv$ diagrams extracted along several position angles
  and slits in NGC\,4151. Contour levels go from $-5\sigma$, $-4\sigma$, $-3\sigma$, $3\sigma$ to $75\sigma$ in steps of $1\sigma=2.8$\,mJy\,beam$^{-1}$. 
The locations of the slits are indicated in
  the intensity map ({\it left}) and the slits are labeled. The $pv$
  diagrams shown are  along (1): the major kinematic axis (PA=22\degr); (2): the northern (PA=0\degr, offset=-7\arcsec; 14\arcsec) and (3): southern (PA=55\degr, offset=-1.5\arcsec; -13\arcsec) gas lanes ({\it from left to right)}.}
\label{pv_diag_lanes} 
\end{figure}

We extracted $pv$ diagrams along three different positions that are
presented in Fig.~\ref{pv_diag_lanes}. The $pv$ diagram along the
major kinematic axis of PA=22\degr\ shows strong emission from the
gas lanes at $\rm +80$\,km\,s$^{-1}$ and $\rm -50$\,km\,s$^{-1}$.
The CO(1-0) line emission shows a velocity gradient indicating that the
gas is participating in circular motion.  Interestingly this gradient is not symmetric with 
respect to the systemic velocity of 995\,km\,s$^{-1}$ from \cite{pedlar_92}, 
it presents an offset of $+15$\,km\,s$^{-1}$. 
 In order to investigate the gas motion within the gas lanes two slits were placed along them
  (number 2 and 3, Fig.~\ref{pv_diag_lanes}).  While the velocity gradient is very low in the
northern lane with a mean velocity of about +80 km\,s$^{-1}$, two
distinct features are present in the southern lane. The eastern part
of the lane shows constant velocities consistent with the systemic
velocity, the western part exhibits a clear velocity gradient from -20
to -60\,km\,s$^{-1}$. 

The CO(1-0) rotation curve of NGC4151 was derived from the velocity
field using the task ROTCUR in GIPSY. This task fits the velocity,
assumed to be circular, within tilted rings. We run ROTCUR for rings
of 2\arcsec\ widths from 3\arcsec\ to 19\arcsec\ radial distances centered on the AGN as
traced by the peak of the 3mm continuum. The region defined by an angle of 5\degr\ 
around the minor axis was excluded for the fit. The inclination and
position angle were fixed at i=21\degr\ and PA=22\degr\ as derived from
the \HI\ data by \cite{mundell_99_2}. The systemic velocity, $v_{sys}$, was allowed to vary
in a first step and found to be at 1010\,km\,s$^{-1}$, i.e.
offset by about +15\,km\,s$^{-1}$ from the value derived by \cite{pedlar_92}. This offset
 is consistent with the offset in the velocity gradient along the major axis, as discussed above.
Indeed, with a systemic velocity of 1010\,km\,s$^{-1}$, the $pv$ diagram along the major kinematic 
axis (PA=22\degr) presents a more symmetric velocity gradient, the southern and northern CO lanes
lying then at -65 and + 65\,km\,s$^{-1}$, respectively.

\begin{figure}
\begin{center}
\includegraphics[width=8cm]{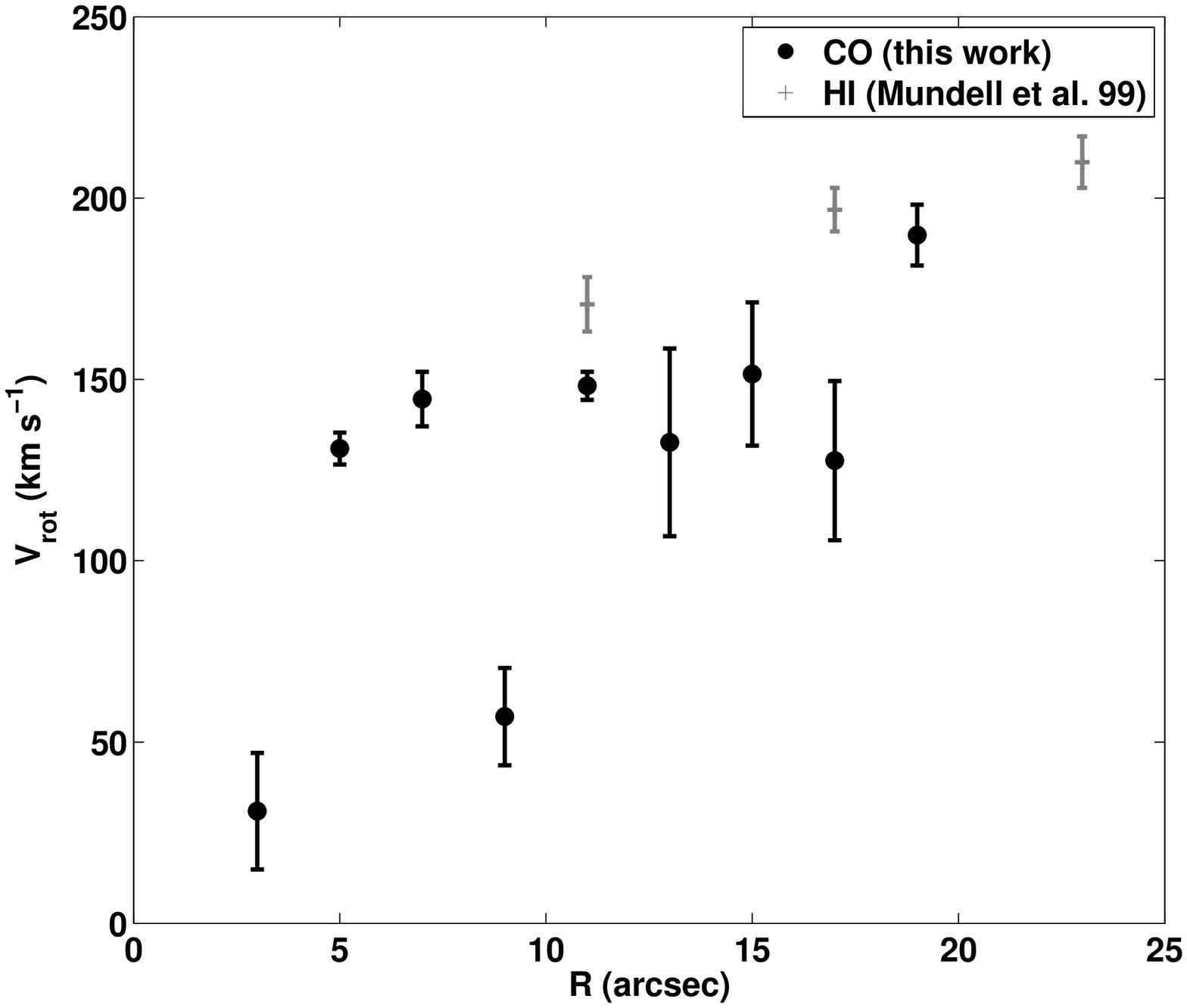} 
\includegraphics[width=8cm]{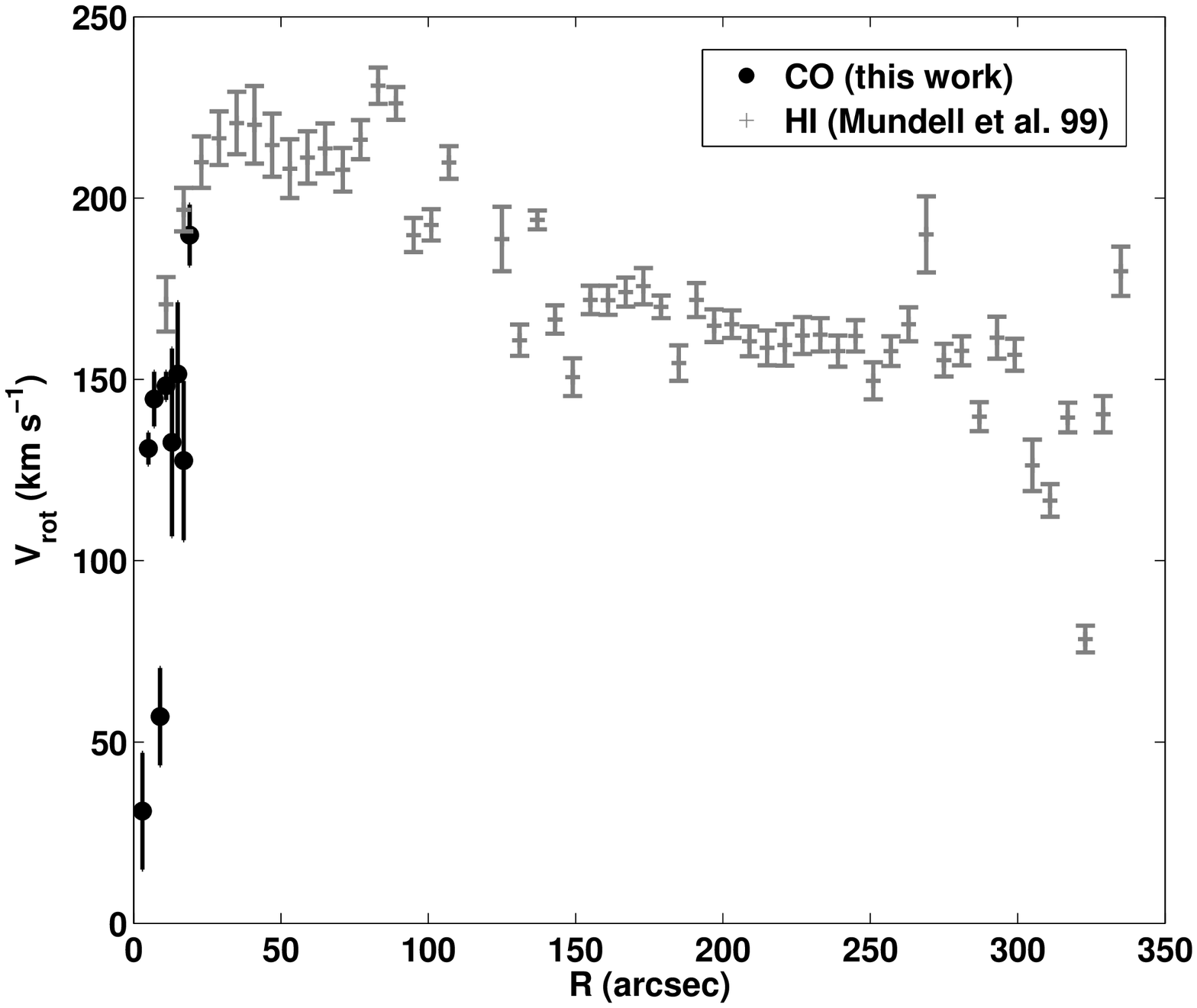} 
\end{center}\caption{ Rotation curve of
  NGC\,4151 derived from the CO(1-0) (black circles) and \HI\ velocity fields
  \citep[gray crosses, from][]{mundell_99_2}. Both panels show the combined CO rotation curve
(derived with v$_{sys}\,=\,1010\,$km\,s$^{-1}$) and HI rotation curve, the left panel showing a zoom on 
the central region. The point at R=9\arcsec\ in CO rotation curve (left panel) corresponds to the north-east part of the southern lane, which has been shown to lie
at the systemic velocity (Fig.~\ref{spectra_mean}) and appears to be at a lower velocity than expected from circular rotation.}
\label{rot_curve} 
\end{figure}

 The resulting rotation curve, extracted with $v_{sys}\,=\,1010$\,km\,s$^{-1}$ is presented in
Fig.~\ref{rot_curve}. The point at 9\arcsec\ presents a velocity  lower than expected for circular rotation. 
This point corresponds to the north-eastern part of the southern lane, which has been shown to have kinematics distinct  
from the west part of the lane (Fig.~\ref{spectra_mean}). The low velocity of the first point at 3\arcsec\ is due to beam smearing.
Comparison to the \HI\ rotation curve \citep{mundell_99_2} shows
a good agreement between the CO and \HI\ rotation curves (Fig.~\ref{rot_curve}, right).

\begin{figure}
\begin{center}
\includegraphics[width=8cm]{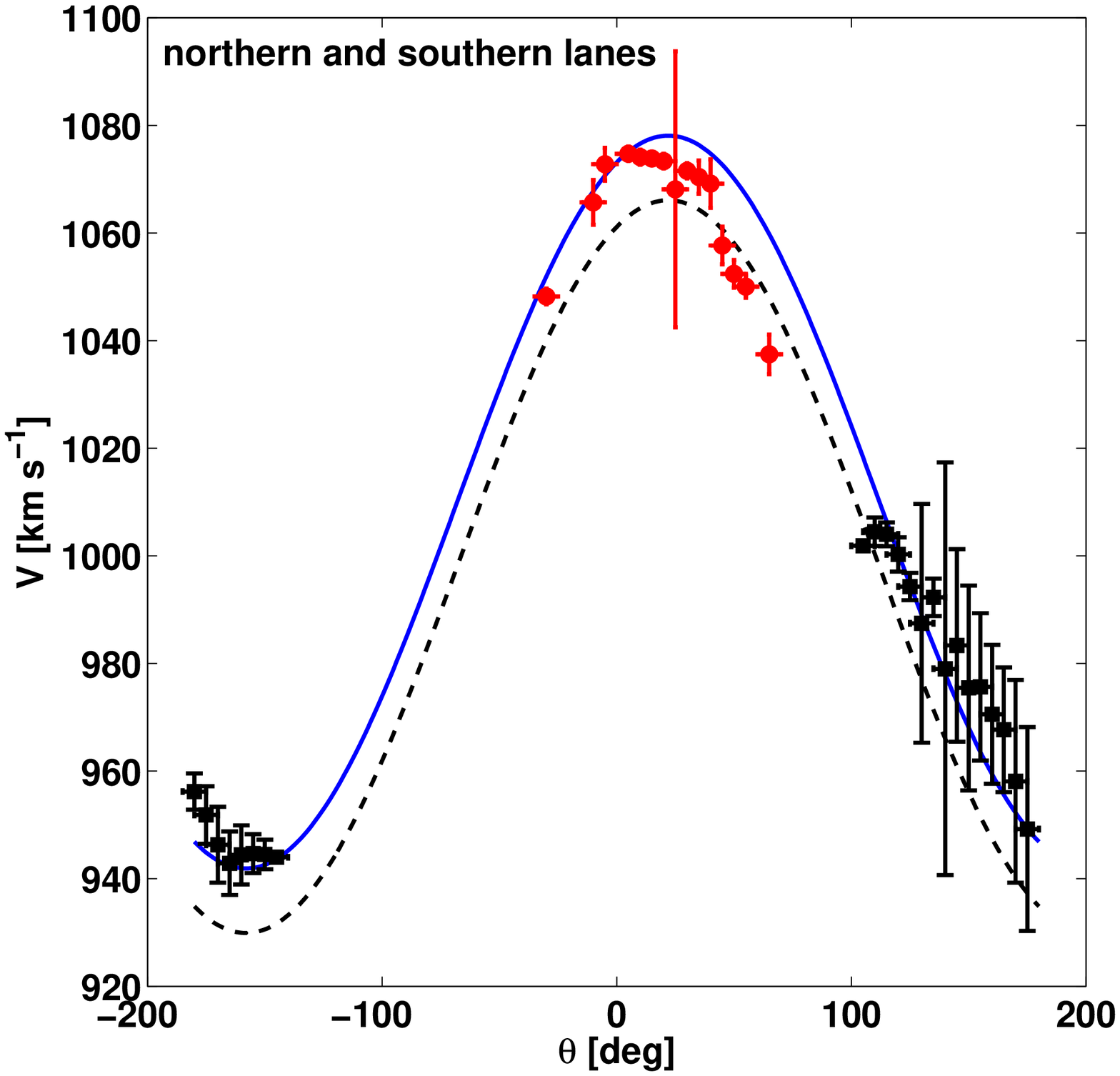} 
\includegraphics[width=8cm]{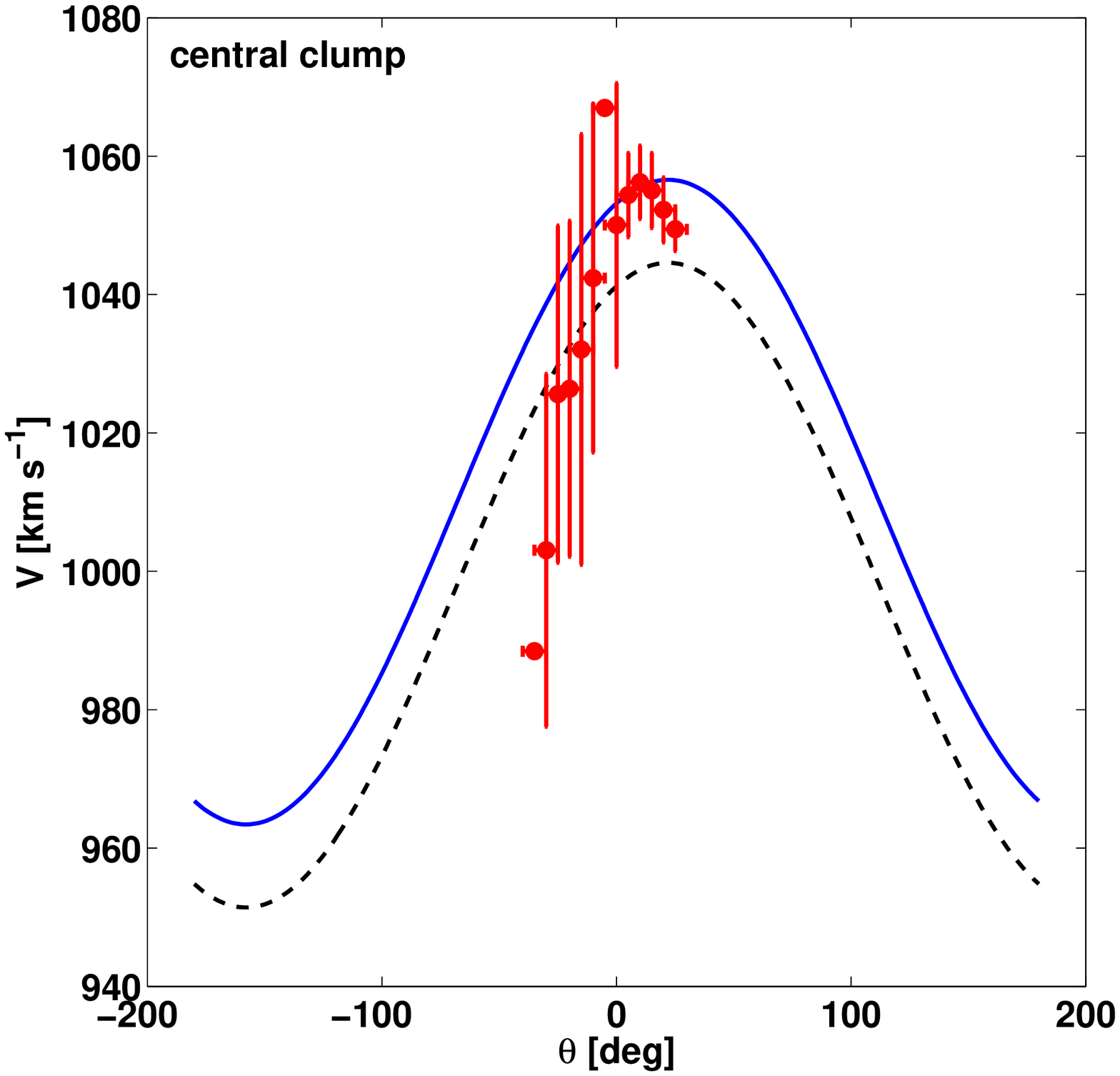} 
\end{center}\caption{{\it Left:} Velocity along the CO lanes, as a function of position
angle $\theta$ around the lanes, from north to east. Red circles (respectively black squares) correspond to the northern lane (respectively southern lane). The two curved lines correspond to  sinusoids of pure circular velocities for an inclined disk with PA=22\degr\ and i=21\degr\  at the major-axis (18\arcsec) of the gas lanes. {\it Right:} Velocity within the central clump, as a function of position angle $\theta$, from north to east (red points). The plain and dashed curves corresponds to sinusoids of pure circular velocities  with disk orientation of PA=22\degr\ and i=21\degr,
at a radius of 5\arcsec, position of the central clump. In both panels, the sinusoids correspond respectively to a systemic velocity of $v_{sys}\,=\,1010\,$km\,s$^{-1}$ (solid blue line), and $v_{sys}\,=\,995\,$km\,s$^{-1}$ (dashed black line). }
\label{vel_vs_theta} 
\end{figure}

Therefore the overall CO kinematics are generally consistent with rotation in the plane of the galaxy. To demonstrate this further, Fig.~\ref{vel_vs_theta} shows velocity as a function of position angle measured directly along the CO arcs. The sinusoidal curves overplotted show the expected variation of velocity along a circular annulus in the plane of the galaxy (i=21$\degr$, PA=22$\degr$) at a radius of 18\arcsec\, corresponding to the semi-major axis of the dust/CO ellipse. Although not exact, the correspondence between the plotted curves is good, ruling out the interpretation that the ellipticity traced by the CO arcs is a result of gas lying in a highly inclined circular disk tilted out of the main galactic plane as well as strong non-circular motions within the disk. The morphology is therefore intrinsically elliptical in the plane of the galaxy, as suggested by the analysis of the \HI\ velocities along the dust arcs \cite[Fig.3 of][]{asif_98} and consistent with predictions of gas flows in bars \citep{athanassoula_92}. The rotation curve in Fig.~\ref{rot_curve} is thus only an approximation to the true kinematics and the offset between the derived CO and HI systemic velocities is a likely indication of disturbed CO gaseous kinematics in these arcs.  

Finally, the velocities within the central clump present larger deviations from circular motions, up to 50\,km\,s$^{-1}$ below the circular velocity at a radius of 5\arcsec (Fig.~\ref{vel_vs_theta}, right). The central clump shows redshifted kinematics (see Fig.~\ref{moments}) significantly inconsistent with those expected for circular motion in this region, close to the minor kinematic axis of the galaxy. These observed deviations are suggestive of streaming motions, however, the uncertainties are too large to derive a more quantitative measure. Very deep CO imaging in the future is required to confirm whether this clump is an isolated cloud or part of a more continuous fluid flow towards the nucleus.

\section{Discussion}
\label{discussion}
\subsection{Role of the  Large-Scale Stellar Bar}
\label{large_scale}

\begin{figure}
\begin{center}
\includegraphics[width=\textwidth]{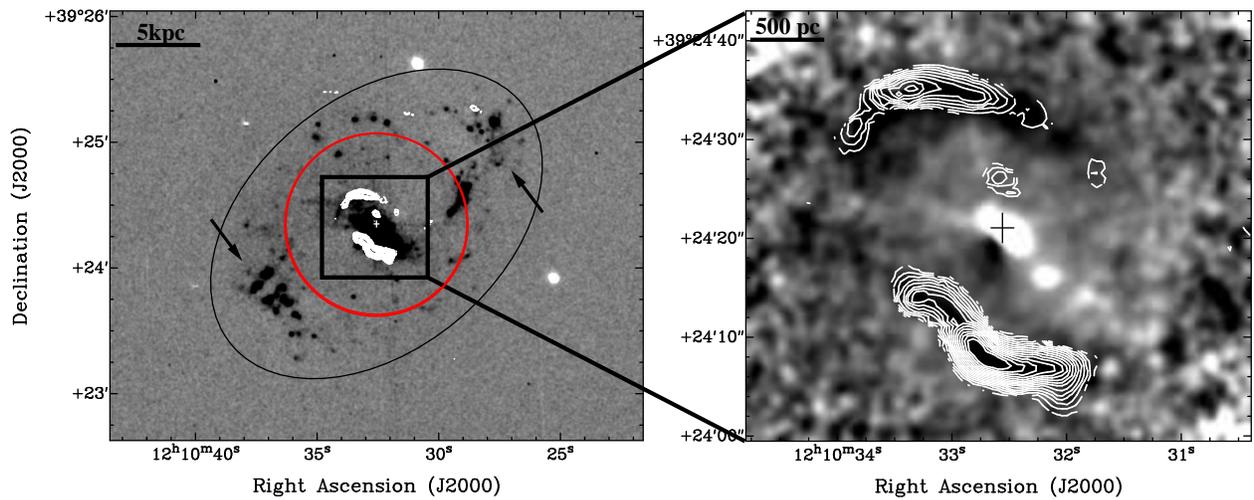} 
\end{center}\caption{ CO(1-0) contours overlaid on the
  H$\alpha$ map \citep[{\it left};][]{knapen_04} and a V-I color map
  \citep[{\it right};][]{asif_98}. The ellipse outlines the
  large-scale stellar bar present in NGC\,4151 and the circle corresponds to the position of the ILR
 as derived by \cite{mundell_99_2}.
 Enhanced H$\alpha$
  emission is present along the leading sides of the bar (indicated by arrows) that connect
  to the CO gas lanes. The V-I map was smoothed to 1.5\arcsec\ resolution
  to enhance the contrast for the dust lanes.  Redder colors are
  presented by darker shading.  The ENLR is evident as the very blue
  elongated shape in the nucleus.  In both panel, the cross marks the position of the
  AGN.}
\label{fig:intensity_dust} 
\end{figure}

In order to analyse the importance of the large-scale stellar bar or
oval for fueling the AGN, we need to establish the link between the
bar and the molecular gas observed in the central 2.5\,kpc. The
detailed \HI\ study of \cite{mundell_99_1} showed that the stellar
bar/oval is unusually gas-rich and that its \HI\ gas dynamics are
consistent with the presence of $x_2$ orbits in an oval bar: sharp velocity
changes across the bright regions close to the leading edges of the
bar are found which can be explained by offset shocks predicted by
simulations \citep{athanassoula_92} and provide direct evidence for 
the presence of these $x_1$ and $x_2$ families of stellar orbits. 
The H$\alpha$ image \citep{knapen_04}
reveals  \HII\  regions along the leading sides of the large-scale oval
coincident with the peaks in the HI surface density. These \HII\ regions are associated 
with the ionized and neutral gas in the streaming shocks \citep{asif_05} that connect 
to the CO gas lanes to form an inner gaseous spiral (Fig.~\ref{fig:intensity_dust}, left). 
\cite{mundell_99_2} derived the resonance 
curves from the \HI\ velocity field and estimated the Inner Lindblad Resonance  (ILR) 
to be at 2.8$\pm$0.6\,kpc. The ILR (red circle in Fig.~\ref{fig:intensity_dust}, left) coincides
with the \HII\ regions, while the CO lanes are well inside the resonance. This picture is 
consistent with the molecular gas tracing a spiral density wave, driven by the large scale 
bar and extending from the ILR down to the center \citep{englmaier_00}. As we do not detect CO all the way down to 
the nucleus, this may argue in favor of the presence of a second ILR that limits the gaseous wave
to exist between the two resonances \citep{englmaier_00}.
 However, the sparse distribution of the CO gas, the lack of \HI\ in the central regions and the non-linear behaviour of gas in a barred potential prevent useful contraints on the presence and location of an inner ILR from being derived directly from the gaseous rotation curves. Full dynamical modelling and numerical simulations constrained by millimetre, radio and optical velocity fields and K-band derived mass models will be presented in a future paper (Dumas et al. in prep.) and may provide insight into location of an inner ILR and its effect on the gas dynamics.

The CO gas lanes also coincide with two dust lanes seen in a V-I color
map \citep{asif_98}, as shown in Fig.~\ref{fig:intensity_dust}, right. The two
red arcs delineate a $11\arcsec\times18\arcsec$ dust ellipse inside
the large oval/bar and also coincide with \HI\ emission \citep{asif_98}. 
The peak column densities in the \HI\ arcs are
$(0.7-1.8)\times10^{21}\,cm^{-2}$ implying an atomic gas mass of $\rm
  M_{HI} \sim (0.13-3.3)\times10^{6}\,M_{\odot}$. Thus the ratio of atomic to molecular
gas is 0.3 in the northern lane and 0.006 in the southern lane, which is in agreement with the
 values of the atomic-to-molecular gas ratio observed in the central regions of spiral
 galaxies \citep{schuster_07,adam_08}, with the southern lane being at the lower end. Moreover, the kinematics of the CO gas lanes and the \HI\ arcs are in good agreement as shown in the previous section. Therefore the morphology and kinematics of the CO northern and southern lanes suggest that the molcular gas lanes  are the continuation of the atomic and dusty ring, along elliptical orbits of the stellar bar. 

It is interesting to note that while the structure of the gas lanes is
very well matched, no obvious
counter-structure is present in the V-I map for the nuclear clump. High resolution HST/ACS imaging shows 
the presence of dust in the central 1.5~kpc, but no clear structure is seen in the dust \citep{bentz_06}.
Connecting the gas flows in the large scale bar with the nuclear dust structures therefore remains a challenge.

\subsection{Cold and warm molecular gas in the innermost region}
\label{COvsH2}

In the central 2\farcs5, excited H$_2$ line emission centered on the nucleus and extended along the minor 
axis of the galaxy ($0\farcs8 \approx 50$\,pc) has been observed by \cite{fernandez_99} and more
recently by \cite{storchi-bergmann_09_1}. The later showed that H$_2$ is thermally excited either
by X-rays from the AGN or by shocks along the accretion flow towards the nucleus. 
This central $H_2$ emission has been interpreted as tracing the reservoir of (cold) gas from
which the super-massive black hole in NGC\,4151 is being fed. In particular, \cite{storchi-bergmann_09_1} 
found a mass of hot H$_2$ gas of $ 240\,M_{\odot}$ and based on the conversion of  warm-to-cold $H_2$ mass  ranging between 10$^{-7}$ and 10$^{-5}$ \citep{dale_05},
they estimated the total mass of  molecular gas (hot plus cold) to be in the range of 2.4$\times$10$^7$-10$^9\,M_{\odot}$. 
 We should easily have detected such an amount of molecular gas. Indeed, the 1$\sigma$ upper limit for the CO(1-0)
emission within the central 3\arcsec\ is 0.033 Jy\,km\,s$^{-1}$, which
translates into a limit for the cold molecular gas mass of 10$^5\,M_{\odot}$, well below the lower limit  expected from the NIR H$_2$ line.

This discrepancy between warm and cold molecular gas mass can be interpreted in two opposite ways. On one hand, it indicates that no significant amount of cold
molecular gas is present close to the nucleus. This implies that the warm molecular gas observed there is not a good
tracer for the general amount of (cold) molecular gas present. Thus the use of a generalized ratio of warm-to-cold H$_2$ could be misleading in the vicinity of AGN, which would explain the large over-estimate of cold molecular gas mass by \cite{storchi-bergmann_09_1}. 
On the other hand, in X-rays dominated regions (XDR), equilibrium molecular abundances may be affected by the X-rays radiation from the AGN. This leads in particular to a depletion of CO abundance with respect to other molecules such as HCN \citep{boger_05}. Indeed many Seyfert galaxies show enhanced central HCN(1-0) emission such as, e.g.  NGC1068 \citep{jackson_93, tacconi_94, helfer_95, usero_04}, NGC1097 \citep{kohno_03}, NGC5194 \citep{kohno_96}, NGC6951 \citep{krips_07}. In particular, the latter study showed that the CO line emission largely underestimates the amount of cold molecular gas in the central region of the Seyfert 2 NGC6951. Therefore,  CO itself may be a poor tracer of cold molecular gas close to an active nucleus.  Under this assumption,  the change of chemistry in the central kpc because of the AGN itself could explain the lack of CO emission inside the dust/gas lanes and the excitation of the NIR H$_2$ would be then driven by the AGN rather than the accretion flow.
Further study of molecular abundances in the central 3\arcsec\ of NGC4151 are needed to quantify the cold molecular mass in the central region of NGC4151 and to explain this discrepancy between (cold) molecular  gas masses inferred from NIR H$_2$ and CO line emission.

\subsection{Implication for AGN fueling}
\label{fueling}

The presence of non-circular motions in the central  clump indicates that fueling of the
AGN or at least a replenishment of the gas reservoir closer to the AGN
might be ongoing. Analysis of near-IR images reveal no evidence for a second inner stellar bar \citep{onken_07}.
 This suggests that the
disturbances of the regular motions might be caused by a transient
event, such as an interaction with a nearby galaxy. Such an interaction has been proposed by  \cite{mundell_99_2} 
to explain the distortion of the large two-arm stellar spiral. NGC4156 and UGC07188 have small projected distances to NGC4151 and therefore may be candidate companions for such a galaxy interaction, although at respectively $v_{sys}$=6750\,km\,s$^{-1}$ and 6850\,km\,s$^{-1}$ the systemic velocities of NGC4156 and UGC07188  are significantly larger than that of NGC4151.

 Our findings are in agreement with recent
results from the NUGA survey
that found no direct indication for ongoing fueling within the central hundred of parsecs 
for most of their targets \citep{garcia-burillo_05,boone_07, lindt_08, casasola_08, haan_09}. 
It is, however, interesting that most NUGA galaxies have a molecular gas
reservoir within the inner kiloparsec with mass between $10^7-10^{10}\msolar$ 
\citep{garcia-burillo_03,combes_04,garcia-burillo_05,krips_05,boone_07,krips_07_1,casasola_08,garcia_burillo_09,combes_09}. It is also interesting to note that most NUGA targets are
either Seyfert type 2 or LINER galaxies.  While gravitational mechanisms, driven e.g. by large scale stellar bars or galaxy interaction, are efficient in transporting gas to the central kpc of galaxies, other processes \citep[e.g. viscous torques,][]{garcia-burillo_05, haan_09} must take over to bring the gas closer to the nucleus and fuel the AGN. Moreover, recent nuclear star formation and AGN activity have been shown to be linked \citep{gonzales_01,riffel_07,davies_07}, and stellar outflows, from starburst on tens of parcsec scales, can also support the nuclear activity \citep{davies_07, hicks_09, schartmann_09}. NGC4151 presents no evidence for a recent nuclear  starburst \citep{sturm_99}. In fact, the star formation rate across the whole galaxy is unusually low given the presence of significant quantities of \HI\ gas \citep{asif_05}, therefore this scenario appears to be unlikely for this particular Seyfert galaxy. Finally, \cite{dumas_07} showed that the ionised gaseous kinematics of Seyferts are more disturbed with respect to the stellar kinematics in the central kiloparsec than those of inactive galaxies, i.e. that kinematic disturbance scales with accretion rate in the inner kiloparsec where dynamical and activity timescales become comparable.

\section{Summary and Conclusions}
\label{conclusion}
We have presented the first interferometric map of the molecular gas
reservoir in NGC\,4151 using PdBI. A total molecular gas mass
of $\rm 4.3\times10^7\,M_{\odot}$ is detected in the central 2.5\,kpc.
Most of the gas is located in two gas lanes at 1\,kpc distance from
the nucleus. A nuclear clump containing about 3\% of the total mass is
found 600\,pc north of the nucleus. However, no CO emission associated
with the AGN has been detected down to a limit of $10^{5}\,M_{\odot}$ for
the central 3'' indicating that estimates of the cold gas reservoir
based on NIR H$_2$ line emission, based on the assumption of the ro-vibrational H$_2$ emission line in thermal equilibrium,
 can lead to severe overestimation of the amount of total gas present.

The morphology and kinematics of the gas lanes are consistent with
being driven by the large-scale stellar bar and also coincide with dust
lanes seen in optical color maps. The detection of non-circular inward motion in the nuclear clump suggests that this cloud may trace gas that will ultimately flow  towards the SMBH.  However, the kinematics of the central clump could not be linked directly to those induced by the large-scale stellar bar. Detailed dynamical modeling of the gravitational potential of the bar and the inner kpc regions is required to combine the molecular and ionised gas and stellar components in a coherent framework and identify the  mechanism responsible for the observed non-circular motions, which is beyond the scope of this paper. 
Thus we speculate here that the potential inflow is caused by a transient
phenomenon and the fueling process of the AGN might be erratic and
intermittent. In agreement with recent studies of
other nearby AGN, these findings suggest that no single (gravitational) mechanism is
responsible for the fueling of nuclear activity.  Finally, in the central kpc where dynamical and activity timescales become comparable, a fine balance may be required between fueling and feedback, but in NGC4151, which has an estimated accretion rate of $\dot{m}=1.3\times10^{-2}M_{\odot}\,\text{yr}^{-1}$ \citep{storchi-bergmann_10}, the upper limit on cold molecular gas in the central 3\arcsec\ ($10^{5}\,M_{\odot}$) given by our CO observations is still sufficient to support the current level of activity of NGC4151 over an AGN lifetime of  $10^7$yr. 

\acknowledgments
We would like to thank the staff on the Plateau de Bure for doing the observations and 
Philippe Salome for his help with the IRAM data reduction.  We also are thankful to Witold Maciejewski for fruitful discussions and to the anonymous referee for useful comments that helped to improve this paper.
CGM acknowledges financial support from the Royal Society and Research Councils U.K.
GD was supported by DFG grants SCH 536/4-1 and SCH 536/4-2 as part of SPP 1177.

Facilities: \facility{IRAM (PdBI)}.

\bibliography{NGC4151_bib}

\clearpage

%%% Fig. 1

\clearpage

%%% Fig. 2

\clearpage

%%% Fig. 3

\clearpage

%%% Fig. 4

\clearpage

%%% Fig.5

%%% Fig. 6

\clearpage

%%% Fig. 7

\clearpage

%%% Fig. 8

\clearpage

%%% Fig. 9

\clearpage

%%% Fig. 10

\clearpage

\clearpage

%%% Tab. 1

\begin{deluxetable}{lrl}
\tabletypesize{\scriptsize}
\tablecaption{General Properties of NCG\,4151\label{properties}}
\tablewidth{0pt}
\tablehead{
\colhead{Property} & 
\colhead{Value} & 
\colhead{Reference} 
}
\startdata
Right ascension (J2000) &   $12^h10^m32.^s58$& \cite{mundell_03}\\
Declination (J2000)& $+39\degr24\arcmin21\farcs06$& \cite{mundell_03}\\
Classification & SAB(rs)ab& RC3 \citep{RC3}\\
AGN Type & Seyfert 1.5& NED\\
Distance& 13.3~Mpc& \cite{mundell_03}\\
Scale& 1\arcsec =65~pc&\\
Systemic Velocity V$_{sys}$  &  995 km s$^{-1}$  & \cite{ulrich_00}\\
Inclination (i)& 21\degr & \cite{mundell_99_2}\\
Position Angle (PA)&22\degr &\cite{mundell_99_2}\\
Bar Extent& 3\arcmin$\times$2\arcmin &\cite{mundell_99_2}\\
Bar PA&130\degr &\cite{mundell_99_2}\\
\enddata

%% Text for table notes should follow after the \enddata but before
%% the \end{deluxetable}. Make sure there is at least one \tablenotemark
%% in the table for each \tablenotetext.

%\tablecomments{}
%\tablenotetext{a}{}
\end{deluxetable}

\clearpage

%%% Tab. 2

\begin{deluxetable}{llrrr}
\tabletypesize{\scriptsize}
\tablecaption{Dataset Parameters\label{beams}}
\tablewidth{0pt}
\tablehead{
\colhead{Dataset} & 
\colhead{Weighting} & 
\colhead{Channel Width} &
\colhead{Clean Beam} &
\colhead{RMS} \\
& &
\colhead{[km s$^{-1}$]} &
\colhead{[\arcsec$\times$\arcsec]} &
\colhead{[mJy beam$^{-1}$]}
}
\startdata
continuum  & natural & 1150.5 & 3.44$\times$2.96 & 0.20 \\
continuum  & uniform &1150.5 & 2.83$\times$2.16 & 0.28 \\ \hline
line-only & natural & 6.5 & 3.44$\times$2.96 & 2.0 \\
line-only & uniform & 6.5 & 2.83$\times$2.16 & 2.8 \\
\enddata

%% Text for table notes should follow after the \enddata but before
%% the \end{deluxetable}. Make sure there is at least one \tablenotemark
%% in the table for each \tablenotetext.

\tablecomments{Beam sizes and noise levels of the PdBI continuum maps and
  line-only data cubes derived at different velocity resolution and
  weightings.}

%\tablenotetext{a}{}
\end{deluxetable}

\clearpage

%%% Tab. 3

\begin{deluxetable}{lrr}
\tabletypesize{\scriptsize}
\tablecaption{CO Fluxes and Molecular Gas Masses\label{mass}}
\tablewidth{0pt}
\tablehead{
\colhead{Component} & 
\colhead{S$_{CO}$dV} & 
\colhead{M$_{mol}$} \\
&
\colhead{[Jy km s$^{-1}$]} &
\colhead{[10$^7$\msolar]} 
}
\startdata
Total & 14.7&4.3\\%2.30\\
Southern lane & 8.7&2.5\\%1.45\\
Northern lane&5.0&1.5\\%0.61\\
Central clump&0.4&0.12\\%0.07\\
Western Clump&0.63&0.18\\%0.05\\
\enddata

%% Text for table notes should follow after the \enddata but before
%% the \end{deluxetable}. Make sure there is at least one \tablenotemark
%% in the table for each \tablenotetext.

\tablecomments{Integrated CO(1-0) line fluxes and corresponding
  molecular gas masses for different components identified in the CO
  distribution. The location and extent of the individual components
  are indicated in Fig.~\ref{components}. The fluxes have been corrected for primary beam attenuation and the masses include 36\% of Helium fraction.}

%\tablenotetext{a}{}
\end{deluxetable}

\end{document}